\documentclass[structabstract]{aa}
\usepackage{graphicx}
\usepackage{natbib}
\usepackage{amsmath,amssymb,graphicx}
\begin{document}
%
%   \title{Accurate measurement of the planetary-system HAT-P-8 parameters}
   \title{A lower radius and mass for the transiting extrasolar planet HAT-P-8\,b}

%   \subtitle{}
\titlerunning{HAT-P-8\,b}

   \author{
          L. Mancini \inst{1,2} %\thanks{}
          \and
          J. Southworth\inst{3}
          \and
          S. Ciceri\inst{1}
          \and
          J. J. Fortney\inst{4}
          \and
          C. V. Morley\inst{4}
          \and
          J. A. Dittmann\inst{5}
          \and
          J. Tregloan-Reed\inst{3}
          \and
          I. Bruni\inst{6}
          \and
          M. Barbieri\inst{7}
          \and
          D. F. Evans\inst{3}
                    \and
          G. D'Ago\inst{2}
                    \and
          N. Nikolov\inst{1}
                    \and
          Th. Henning\inst{1}}
       \institute{Max Planck Institute for Astronomy, K\"{o}nigstuhl 17, 69117 -- Heidelberg, Germany \\
             \email{mancini@mpia.de}
         \and
    Department of Physics, University of Salerno, Via Ponte Don Melillo, 84084 -- Fisciano (SA), Italy
         \and
    Astrophysics Group, Keele University, Staffordshire, ST5 5BG, UK
        \and
    Department of Astronomy \& Astrophysics, University of California, Santa Cruz, CA 95064, USA
        \and
    Harvard-Smithsonian Center for Astrophysics, 60 Garden Street, MS 10 Cambridge, MA 02138, USA
        \and
    INAF -- Osservatorio Astronomico di Bologna, Via Ranzani 1, 40127 -- Bologna, Italy %
        \and
    INAF -- Osservatorio Astronomico di Padova, Vicolo Osservatorio 5, 35122 -- Padova, Italy}

   \date{Received ; Accepted}
% \abstract{}{}{}{}{}
% 5 {} token are mandatory
  \abstract
{The extrasolar planet HAT-P-8\,b was thought to be one of the
more inflated transiting hot Jupiters.}
{By using new and existing photometric data, we computed precise
estimates of the physical properties of the system.}
{We present photometric observations comprising eleven light
curves covering six transit events, obtained using five
medium-class telescopes and telescope-defocussing technique. One
transit was simultaneously obtained through four optical filters,
and two transits were followed contemporaneously from two
observatories. We modelled these and seven published datasets
using the {{\sc jktebop}} code. The physical parameters of the
system were obtained from these results and from published
spectroscopic measurements. In addition, we investigated the
theoretically-predicted variation of the apparent planetary radius
as a function of wavelength, covering the range $330$--$960$\,nm.}
{We find that HAT-P-8\,b has a significantly lower radius ($1.321
\pm 0.037 \, R_{\mathrm{Jup}}$) and mass ($1.275 \pm 0.053 \,
M_{\mathrm{Jup}}$) compared to previous estimates
($1.50_{-0.06}^{+0.08} \, R_{\mathrm{Jup}}$ and
$1.52_{-0.16}^{+0.18} \, M_{\mathrm{Jup}}$ respectively). We also
detect a radius variation in the optical bands that, when compared
with synthetic spectra of the planet, may indicate the presence of
a strong optical absorber, perhaps TiO and VO gases, near the
terminator of HAT-P-8\,b.}
{These new results imply that HAT-P-8\,b is not significantly
inflated, and that its position in the planetary mass--radius
diagram is congruent with those of many other transiting
extrasolar planets.}


\maketitle

%
%%%%%%%%%%%%%%%%%%%%%%%%%%%%%%%%%%%%%%%%%%%%%%%%%%%%%%
\section{Introduction}
\label{sec_1}
%%%%%%%%%%%%%%%%%%%%%%%%%%%%%%%%%%%%%%%%%%%%%%%%%%%%%%
Transiting extrasolar planetary systems are of interest and
importance, because precise measurements of their physical
properties can be achieved using spectroscopic and photometric
observations. Atomic and molecular absorption within the
atmosphere of transiting extrasolar planets (TEPs) can also be
investigated through transmission spectroscopy (e.g.\
\citealp{swain2008,sing2009,fossati2010}) and simultaneous
multi-colour photometry (e.g.\
\citealp{ballester2007,southworth2012b}) of the transits.
High-quality photometric observations not only enable the
measurement of the masses and radii of TEPs to accuracies of a few
percent (e.g. \citealp{torres2008,southworth09}), but also the
detection of transit anomalies due to stellar pulsations
\citep{cameron2010}, tidal distortion \citep{li2010,leconte2011},
additional bodies (moons, planets)
\citep{kipping2009,tusnski2011}, gravity darkening
\citep{barnes2009,szabo2011} and star spots
\citep{pont2007,rabus2009,desert2011}. Besides the
Rossiter-McLaughlin effect \citep{queloz00,gaudi2007}, the
photometric follow-up on consecutive/close nights of transits of
planets over parent-star starspots represents another fascinating
method \citep{sanchis2011,sanchis2011b,sanchis2012,tregloan2012}
to measure the sky-projected spin-orbit alignment.

The increasing number of TEPs discovered every year is
progressively revealing a remarkabel diversity. The improving
statistical weight of this sample is useful for establishing the
correct theoretical framework of planet formation and evolution.
Accurate estimates of the planet properties (mass, radius, orbital
semi-major axis, etc.) are vital for this purpose, and photometric
follow-up of known TEPs can dramatically improve our knowledge of
the planet's characteristics.
% (e.g.\ \citealp{southworth2011}).

HAT-P-8\,b is a transiting hot Jupiter found by the HATNet team
\citep{latham2009}, orbiting with a period of $\sim3.07$ days
around a star of spectral type F8 \citep{jones2010} or F5
\citep{bergfors2012}. At the time of its discovery it was labelled
as one of the most inflated transiting giant planets, with a
measured mass and radius of $M_{\rm b} =
1.52^{+0.18}_{-0.16}\,M_{\mathrm{Jup}}$ and $R_{\rm b} =
1.50^{+0.08}_{-0.06}\,R_{\mathrm{Jup}}$, respectively. These
values differ by 2--3$\sigma$ from the theoretical predictions of
\citet{fortney2007}.

The Rossiter-McLaughlin effect has been detected in the HAT-P-8
system using radial velocity observations from the SOPHIE and FIES
spectrographs. \citet{simpson2011} found a sky-projected orbital
obliquity of $\lambda = -9.7^{+9.0}_{-7.7}$\,degrees and
\citet{moutou2011} found $\lambda = -17^{+9.2}_{-11.5}$\,degrees;
both values are consistent with alignment between the orbital axis
of the planet and the rotational axis of the star. The two studies
between them suggested lower values for $M_{\rm b}$ and $R_{\rm
b}$, but neither calculated the physical properties of the system.

\citet{bergfors2012} have found a faint companion to the HAT-P-8
system using \emph{lucky imaging} observations with the AstraLux
Norte instrument at the Calar Alto 2.2\,m telescope. The
companion, a likely M2-4 dwarf, is at an angular distance of
$1.027\pm 0.011$\,arcsec and is fainter in the SDSS $i^\prime$ and
$z^\prime$ passbands by $\Delta i^\prime = 7.34 \pm 0.10$\,mag and
$\Delta z^\prime = 6.68 \pm 0.07$\,mag. The faintness of this star
means that it has a negligible effect on optical observations of
HAT-P-8.

In this work we present eleven new follow-up light curves six
transits in the HAT-P-8 system, obtained using five 1.2--2.5\,m
telescopes. We augment these data with previously published
observations of seven transits, and measure the physical
properties of the system. We find a substantially lower mass and
radius for the planet, removing its outlier status and relegating
it to a more well-populated part of the planetary mass-radius
diagram.

Our paper is structured as follows. In Sect.\,\ref{sec_2} we
describe the observations and data reduction. In
Sect.\,\ref{sec_3} we analyse the data, and in Sect.\,\ref{sec_4}
we obtain refined orbital ephemerides and physical properties of
the HAT-P-8 system. In Sect.\,\ref{sec_5} we investigate the
variation of the planetary radius as function of wavelength.
Several anomalies detected in the light curves are discussed in
Sect.\,\ref{sec_6}, whereas in Sect.\,\ref{sec_7} we summarize the
results and draw our conclusions.

%%%%%%%%%%%%%%%%%%%%%%%%%%%%%%%%%%%%%%%%%%%%%%%%%%%%%%
\section{Observations and data reduction}
\label{sec_2}
%%%%%%%%%%%%%%%%%%%%%%%%%%%%%%%%%%%%%%%%%%%%%%%%%%%%%%

% Figure 01
\begin{figure*}%
\centering
\includegraphics[width=18.cm]{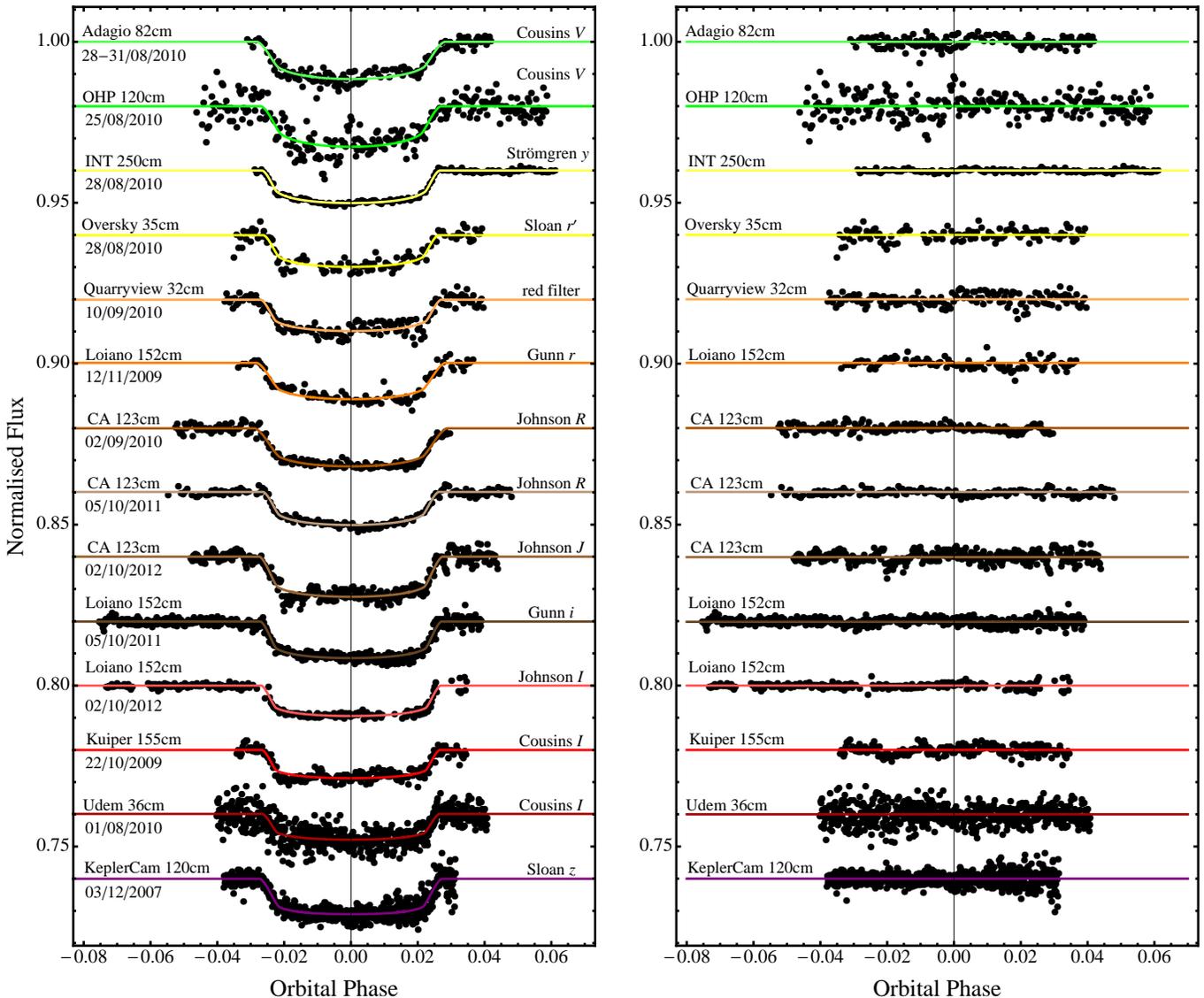}
\caption{Phased light curves of HAT-P-8 compared to the best {\sc
jktebop} fits using the quadratic LD law (left-hand panel). They
are ordered according to central wavelength of the filter used.
The residuals of the fits are plotted in the right-hand panel,
offset to bring them into the same relative position as the
corresponding best fit in the left-hand panel.} \label{Fig_01}
\end{figure*}

Six transits of HAT-P-8\,b were monitored by five different
telescopes between 2009 and 2012. Two transits were followed
simultaneously by two of the telescopes, and one was
simultaneously observed through four optical filters. Except the
first, all our transit observations were performed with the
telescope defocussing method, in order to minimise the effect of
Poisson, scintillation and flat-fielding noise
\citep{southworth2009a}. We used autoguiding during all the
observations, and the stars incurred a drift of 10 pixels or less
over each observing sequence. The corresponding night logs are
reported in Table\,\ref{Table_1}, and the differential photometry
is tabulated in Table\,\ref{Table_2} and plotted in
Figs.\,\ref{Fig_01} and\,\ref{Fig_02}.

% Table 1
\begin{table*}
\caption{Details of the observations presented in this work.
$N_{\mathrm{obs}}$ is the number of observations, Moon is the
fractional illumination of the Moon at the midpoint of the
transit, and $t_{\rm exp}$ is the exposure time in seconds. The
aperture sizes are the radii in pixels of the software apertures
for the star, inner sky and outer sky, respectively. Scatter is
the r.m.s.\ scatter of the data versus a fitted model, in mmag.
Target mean PSF area in px$^{2}$ is also reported for each
dataset. Times and dates are in UT. $\beta$ is the ratio between
the noise levels due to Poisson noise and to combined Poisson and
red noise.}

\label{Table_1}%
\centering %
\tiny %
\begin{tabular}{lcccrrcccccc}
% columns
\hline\hline
Telescope & Date & Start/End times & $N_{\mathrm{obs}}$ & $t_{\rm exp}$ & Filter & Airmass & Moon & Aperture & Scatter & PSF & $\beta$ \\
\hline %
Kuiper   & 2009 10 22 & $04:03 \rightarrow 09:10$ & 1083 &   5 & Cousins $I$       & $1.00 \rightarrow 1.9$                   & $25\%$ & 10, 20, 25 & 1.54 &  \\
Loiano   & 2009 11 12 & $16:52 \rightarrow 22:09$ &  103 & 118 & Gunn $r$          & $1.02 \rightarrow 1.01 \rightarrow 1.55$ & $23\%$ & 25, 45, 60 & 1.15 & 1963 & 1.4 \\
INT      & 2010 08 28 & $20:44 \rightarrow 04:02$ &  131 & 140 & Str\"{o}mgren $y$ & $2.04 \rightarrow 1.00 \rightarrow 1.18$ & $84\%$ & 45, 65, 90 & 0.47 & 3217 & 1.7  \\
Loiano   & 2011 10 05 & $18:01 \rightarrow 02:29$ &  469 &  50 & Gunn $i$          & $1.12 \rightarrow 1.01 \rightarrow 1.45$ & $61\%$ & 19, 40, 60 & 1.15 & 908  & 1.4 \\
CA 1.23m & 2011 10 05 & $19:27 \rightarrow 03:07$ &  166 &  70 & Johnson $R$       & $1.08 \rightarrow 1.05 \rightarrow 1.54$ & $61\%$ & 26, 38, 55 & 0.83 & 1662 & 1.0 \\
CA 2.2m  & 2012 08 26 & $20:31 \rightarrow 04:48$ &  160 &  80 & sdss $u$          & $1.52 \rightarrow 1.00 \rightarrow 1.55$ & $77\%$ & 25, 55, 75 & 2.37 & 2827 & 1.0 \\
CA 2.2m  & 2012 08 26 & $20:31 \rightarrow 04:48$ &  157 &  80 & Gunn $g$          & $1.52 \rightarrow 1.00 \rightarrow 1.55$ & $77\%$ & 30, 50, 70 & 0.97 & 2463 & 1.2 \\
CA 2.2m  & 2012 08 26 & $20:31 \rightarrow 04:48$ &  159 &  80 & Gunn $r$          & $1.52 \rightarrow 1.00 \rightarrow 1.55$ & $77\%$ & 30, 60, 90 & 0.71 & 2642 & 1.5 \\
CA 2.2m  & 2012 08 26 & $20:31 \rightarrow 04:48$ &  162 &  80 & Gunn $z$          & $1.52 \rightarrow 1.00 \rightarrow 1.55$ & $77\%$ & 28, 55, 80 & 0.92 & 1662 & 1.1 \\
Loiano   & 2012 10 02 & $17:45 \rightarrow 02:32$ &  232 &  70 & Johnson $I$       & $1.03 \rightarrow 1.00 \rightarrow 1.92$ & $92\%$ & 23, 40, 65 & 0.79 & 2521 & 1.0 \\
CA 1.23m & 2012 10 02 & $20:07 \rightarrow 03:35$ &  373 &  50 & Johnson $I$       & $1.11 \rightarrow 1.00 \rightarrow 2.16$ & $92\%$ & 35, 60, 90 & 1.69 & 2290 & 1.8 \\
\hline %
\end{tabular}
\end{table*}

%%%%%%%%%%%%%%%%%%%%%%%%%%%%%%%%%%%%%%%%%%%%%%%%%%%%%%
\subsection{Kuiper 1.55\,m telescope }
\label{sec_2.1}
%%%%%%%%%%%%%%%%%%%%%%%%%%%%%%%%%%%%%%%%%%%%%%%%%%%%%%
We observed one transit of HAT-P-8 in October 2009 using the
University of Arizona's 1.55\,m Kuiper telescope on Mt.\ Bigelow,
Arizona. We used the Mont4k CCD, binned 3$\times$3 to
0.43$^{\prime\prime}$/pixel, for a total field of view (FOV) of
$9.7^{\prime} \times 9.7^{\prime}$, and an Arizona-$I$
filter\footnote{The transmission curve for this filter is shown
at:
james.as.arizona.edu/{\textasciitilde{psmith}}/61inch/FILTERS/harris.jpg}.
The observations were conducted with autoguiding and the telescope
focussed. Due to the bright nature of the star we used 5\,s
exposure times, which resulted in an observing cadence of
$\approx$$30$\,s.

Systematic effects were minimised by autoguiding, keeping star
wander to less than 5 pixels (2.15\arcsec) over the course of the
night. The resulting images were bias-subtracted, flat-fielded,
and bad pixel-cleaned in the usual manner. Aperture photometry was
performed using an {\sc idl}\footnote{The acronym {\sc idl} stands
for Interactive Data Language and is a trademark of ITT Visual
Information Solutions. For further details see {\tt
http://www.ittvis.com/ProductServices/IDL.aspx}.} pipeline
utilising the {\sc find} and {\sc aper} \citep{stetson1987} tasks
available in the NASA Astronomy User's Library\footnote{The IDL
Astronomy User's Library (ASTROLIB) is available at
http://idlastro.gsfc.nasa.gov/}. The size of the aperture was
chosen to minimise scatter in the data and was 10 pixels
(4.3\arcsec) in radius. Several combinations of reference stars
were considered, and the one which gave the lowest scatter in the
final light curve was adopted. The 1083 original datapoints were
binned to yield 216 final datapoints.

%%%%%%%%%%%%%%%%%%%%%%%%%%%%%%%%%%%%%%%%%%%%%%%%%%%%%%
\subsection{Cassini 1.52\,m telescope}
\label{sec_2.2}
%%%%%%%%%%%%%%%%%%%%%%%%%%%%%%%%%%%%%%%%%%%%%%%%%%%%%%
One transit of HAT-P-8 was observed in November 2009, two in
October 2011 and one in October 2012, using the 1.52\,m Cassini
Telescope at the Astronomical Observatory of Bologna in Loiano
(Italy). We have previously used this telescope several times to
observe planetary transits \citep[e.g.][]{southworth2012a}, with
the BFOSC (Bologna Faint Object Spectrograph \& Camera) instrument
operated in imaging mode.

The CCD was used unbinned, giving a plate scale of
$0.58^{\prime\prime}/\rm{pixel}$, for a total FOV of $13^{\prime}
\times 12.6^{\prime}$, and the telescope was autoguided and
defocussed. The first transit was observed through a Gunn $r$
filter, the two 2011 transits through a Gunn $i$ filter, and the
last one through a Johnson $I$ filter. The first 2011 transit
suffered from systematic noise due to a bad pixel in the aperture
of the target star, so we did not use these data in our analysis.
The transit of 2012 was disturbed by clouds, which affected the
photometry particularly at the end of the transit. We removed the
points compromised by the clouds.

The observations were analysed using the {\sc idl} pipeline from
\citet{southworth2009a}. The images were debiased and flat-fielded
using standard methods, then subjected to aperture photometry
using the {\sc aper} task. Pointing variations were followed by
cross-correlating each image against a reference image. We chose
the aperture sizes and comparison stars which yielded the lowest
scatter in the final differential-photometry light curve. The
relative weights of the comparison stars were optimised
simultaneously with fitting a second-order polynomial to the
outside-transit observations in order to normalise them to unit
flux.

%%%%%%%%%%%%%%%%%%%%%%%%%%%%%%%%%%%%%%%%%%%%%%%%%%%%%%
\subsection{Isaac Newton Telescope}
\label{sec_2.3}
%%%%%%%%%%%%%%%%%%%%%%%%%%%%%%%%%%%%%%%%%%%%%%%%%%%%%%
One transit of HAT-P-8 was monitored using the 2.5\,m Isaac Newton
Telescope (INT), La Palma (Spain), equipped with the Wide Field
Camera (WFC) at prime focus. We used only one of the four CCDs,
unbinned and with a plate scale of
$0.33^{\prime\prime}/\rm{pixel}$, for a total FOV of
$12.6^{\prime} \times 11.3^{\prime}$. The telescope was defocussed
and autoguided, and the observations were obtained through a
Str\"omgren $y$ filter. A few datapoints at the start of the
observing sequence were rejected as they are affected by
systematic noise (due probably to high airmass $>$2). The
observations were reduced in the same way as those from the
Cassini Telescope (Sect.\,\ref{sec_2.2}).

% Figure 02
\begin{figure}%
\centering
\includegraphics[width=9.cm]{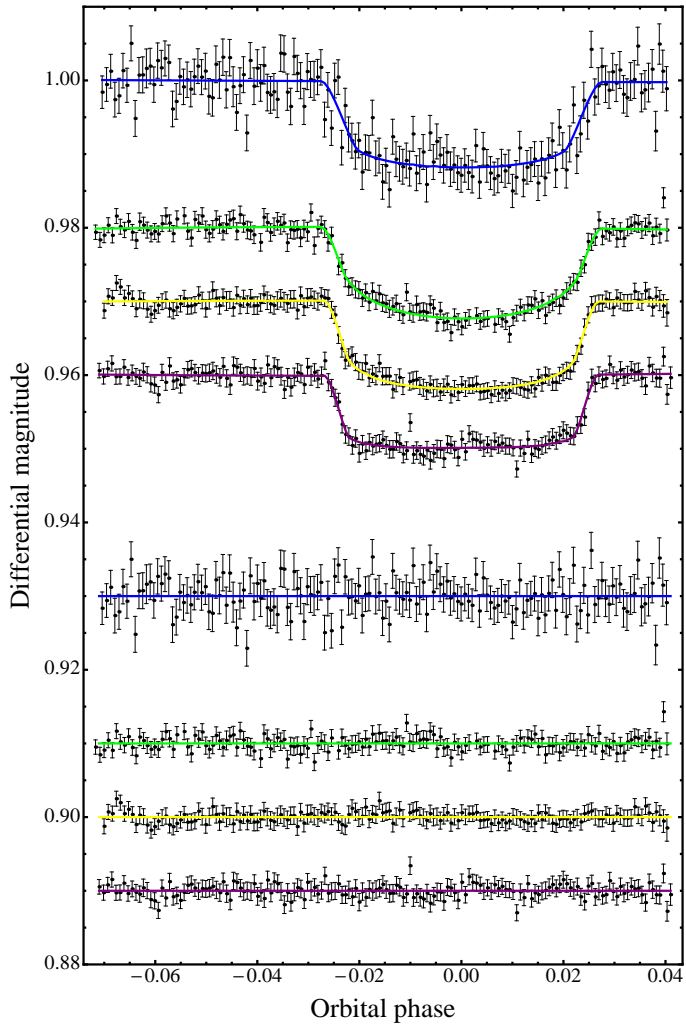}
\caption{Phased BUSCA light curves of HAT-P-8 compared to the best
{\sc jktebop} fits using the quadratic LD law (left-hand panel).
They are ordered according to central wavelength of the filter
used (sdss $u$, Gunn $g$, Gunn $r$, Gunn $z$). The residuals of
the fits are plotted at the base of the figure, offset from zero.} %
\label{Fig_02}
\end{figure}

%%%%%%%%%%%%%%%%%%%%%%%%%%%%%%%%%%%%%%%%%%%%%%%%%%%%%%
\subsection{Calar Alto 1.23\,m telescope}
\label{sec_2.4}
%%%%%%%%%%%%%%%%%%%%%%%%%%%%%%%%%%%%%%%%%%%%%%%%%%%%%%
Three transits of HAT-P-8 b, two in October 2011 and one in
October 2012, were obtained using the 1.23\,m telescope at the
German-Spanish Calar Alto Observatory (CAHA) near Almer\'{i}a,
Spain. Autoguiding was used. During the 2011 observations, we used
the 2k$\times$2k SITE\#2b otical CCD, which had\footnote{This CCD
has been decommissioned.} a FOV of $16^{\prime} \times
16^{\prime}$ and a pixel scale of $0.5^{\prime\prime}$ per pixel.
We defocussed the telescope, read out only a small window in order
to limit the dead time between exposures, and observed through a
Johnson $R$ filter. The first transit was incomplete due to
clouds, so was not included in our analysis. The 2012 transit was
obtained through a Johnson $I$ filter with the new DLR-MKIII
camera, which is equipped with an e2v CCD231-84-NIMO-BI-DD sensor
with 4k$\times$4k pixels and a FOV of $21^{\prime} \times
21^{\prime}$ at $0.3^{\prime\prime}$ per pixel. Unfortunately, the
operator did not included appropriate reference stars in the field
of view, and the scatter of the resulting light curve is higher
than those of 2011. All the observations were reduced as for the
Cassini Telescope (Sect.\,\ref{sec_2.2}).

% Table 2
\begin{table}
\caption{Excerpts of the light curves of HAT-P-8: this table will
be made available at the CDS. A portion is shown here for guidance
regarding its form and content.} %
\label{Table_2}%
\centering %
\tiny
\begin{tabular}{llcrc}
% columns
\hline\hline
Telescope    & Filter & BJD (TDB) & Diff. mag. & Uncertainty  \\
\hline %
Loiano       & $r$      & 2455148.210781 &  0.0009980  & 0.0018456 \\
\vspace{0.1cm}  %
Loiano       & $r$      & 2455148.214867 &  0.0000050  & 0.0017916 \\
Loiano       & $i$      & 2455840.256418 &  0.0000785  & 0.0016988 \\
\vspace{0.1cm}  %
Loiano       & $i$      & 2455840.258316 &  0.0006256  & 0.0016925 \\
CAHA         & $R$      & 2455840.319832 &  0.0007127  & 0.0013972 \\
\vspace{0.1cm}  %
CAHA         & $R$      & 2455840.329288 &  -0.0019752 & 0.0013798 \\
INT          & $y$      & 2455437.412356 &  0.0052976  & 0.0009779 \\
\vspace{0.1cm}  %
INT          & $y$      & 2455437.414575 &  0.0056092  & 0.0009998 \\
Kuiper       & $I$      & 2455126.66906  &  -0.99507   & $-$ \\
\vspace{0.1cm}  %
Kuiper       & $I$      & 2455126.66930  &  -0.99959   & $-$ \\
\hline %
\end{tabular}
\end{table}

%%%%%%%%%%%%%%%%%%%%%%%%%%%%%%%%%%%%%%%%%%%%%%%%%%%%%%
\subsection{Calar Alto 2.2\,m telescope}
\label{sec_2.5}
%%%%%%%%%%%%%%%%%%%%%%%%%%%%%%%%%%%%%%%%%%%%%%%%%%%%%%
We observed one full transit of HAT-P-8 on the night of 2012
August 26, using the 2.2\,m telescope and BUSCA imager at CAHA.
BUSCA is designed for simultaneous four-colour photometry: the
light is split into four wavelength bands from UV to visual IR
using three dichroics. In the four corresponding focal planes the
same area of the sky is imaged onto 4k$\times$4k $15\mu$m pixel
CCDs. For our observations we chose to have the SDSS $u$ filter in
the bluest arm and standard Calar Alto Gunn $g$, $r$ and $z$
filters in the other three arms. This choice led to a reduced
field of view (from $12^{\prime} \times 12^{\prime}$ to a circle
of $6^{\prime}$ in diameter), but had the advantage of a much
better throughput in $grz$ compared to the default Str\"{o}mgren
filters. We defocussed BUSCA in such a way as to have as much
signal as possible in the $u$ band whilst remaining in the linear
regime in the other passbands. The CCDs were binned $2 \times 2$
to shorten the readout time. The autoguider was operated in-focus.
The observations were reduced in the same way as those from the
Cassini Telescope (Sect.\,\ref{sec_2.2}) and the resulting light
curves are plotted in Fig.\,\ref{Fig_02}.

%%%%%%%%%%%%%%%%%%%%%%%%%%%%%%%%%%%%%%%%%%%%%%%%%%%%%%
\section{Light curve analysis}
\label{sec_3}
%%%%%%%%%%%%%%%%%%%%%%%%%%%%%%%%%%%%%%%%%%%%%%%%%%%%%%
Sect.\,\ref{sec_2} introduced eleven transit light curves which
were suitable for detailed analysis. These were each fitted using
the {\sc jktebop}\footnote{\textsc{jktebop} is written in
FORTRAN77 and the source code is available at
http://www.astro.keele.ac.uk/{\textasciitilde{jkt}}/} code
\citep{southworth2004}, which represents the star and planet as
biaxial spheroids. The primary parameters of the fit were the
orbital inclination, $i$, the sum and ratio of the fractional
radii of the star and planet, $r_{\mathrm{A}}+r_{\mathrm{b}}$ and
$k = r_{\mathrm{b}}/r_{\mathrm{A}}$, defined as $r_{\mathrm{A}} =
R_{\mathrm{A}}/a$ and $r_{\mathrm{b}} = R_{\mathrm{b}}/a$, and
transit midpoint, $T_0$. The orbital semimajor axis is $a$ and,
$R_{\mathrm{A}}$ and $R_{\mathrm{b}}$ are the absolute radii of
the two celestial bodies.

Once a fit was available for each dataset, we rescaled the
errorbars to give a reduced $\chi^{2}$ of $\chi_{\nu}^{2}=1$. This
step is necessary because the {\sc aper} aperture photometry
procedure has a tendency to underestimate the measurement errors.
Then, in order to take systematic noise into account, we inflated
the errorbars further using the $\beta$ approach (e.g.\
\citealt{gillon2006, winn2008, winn2009, gibson2008,
nikolov2012}). We calculated $\beta$ values for between two and
ten datapoints for each light cuurve, and adopted the largest
$\beta$ value.

Limb darkening (LD) was accounted for using a quadratic law. The
linear LD coefficient was fitted, whereas the non-linear one was
fixed at a theoretically predicted value \citep{claret2004}, but
perturbed by $\pm$0.1 during the process of error estimating. The
atmospheric parameters of the star assumed for deriving the
limb-darkening coefficients were: $T_{\mathrm{eff}}=6130$, $\log
g=4.15$, [Fe/H]=+0.01, $V_{\mathrm{micro}}=2\,$km s$^{-1}$.
Uncertainties in the fitted parameters from each solution were
calculated in two ways: from 1000 Monte Carlo simulations and with
a residual-permutation algorithm (see \citealt{southworth08}). The
larger of the two possible error bars was retained in each case.
The light curves and their best-fitting models are shown in
Figs.\,\ref{Fig_01} and\,\ref{Fig_02}, whereas the parameters of
each fit are reported in Table\,\ref{Table_3}.

We also attempted to fit all light curves simultaneously using the
{\sc tap} \citet{gazak} code, but were unsuccessful. Analyses of
individual light curves, however, yielded similar results and
uncertainties as to those from {\sc jktebop}.

% Table 3
\begin{table*}
\caption{Parameters of the {\sc jktebop} fits to the light curves
of HAT-P-8. The final parameters, given in bold, are the weighted
means of the results for the 11 datasets. Results from
theliterature are included at the base of the table for
comparison.}

\label{Table_3} %
\centering %
\tiny
\begin{tabular}{llccccc}
% columns
\hline\hline
Source & Filter & $r_{\mathrm{A}}+r_{\mathrm{b}}$ & $k$ & $i^{\circ}$ & $r_{\mathrm{A}}$ & $r_{\mathrm{b}}$  \\
\hline %
Loiano \#1        & Gunn $r$           & $0.177  \pm 0.013  $ & $0.0921  \pm 0.0034 $ & $87.3 \pm 2.3$ & $0.162  \pm 0.012 $ & $0.0149  \pm 0.0014 $ \\
Loiano \#2        & Gunn $i$           & $0.1695 \pm 0.0030 $ & $0.0943  \pm 0.0022 $ & $89.2 \pm 1.0$ & $0.1460 \pm 0.0026$ & $0.01460 \pm 0.00046$ \\
Loiano \#3        & Johnson $I$        & $0.1674 \pm 0.0035 $ & $0.0884  \pm 0.0010 $ & $88.5 \pm 1.0$ & $0.1538 \pm 0.0032$ & $0.01360 \pm 0.00034$ \\
CA 1.23\,m \#1    & Johnson $R$        & $0.218  \pm 0.012  $ & $0.1006  \pm 0.0021 $ & $82.7 \pm 1.1$ & $0.198  \pm 0.010 $ & $0.0199  \pm 0.0012 $ \\
CA 1.23\,m \#2    & Johnson $R$        & $0.1685 \pm 0.0061 $ & $0.0888  \pm 0.0032 $ & $88.7 \pm 1.6$ & $0.1548 \pm 0.0058$ & $0.01374 \pm 0.00084 $ \\
CA 1.23\,m \#3    & Johnson $I$        & $0.208  \pm 0.018  $ & $0.1056  \pm 0.0042 $ & $83.3 \pm 1.6$ & $0.188  \pm 0.016 $ & $0.0199  \pm 0.0018 $ \\
CA 2.2\,m         & sdss $u$           & $0.199  \pm 0.024  $ & $0.0993  \pm 0.0039 $ & $84.1 \pm 2.5$ & $0.181  \pm 0.021 $ & $0.0180  \pm 0.0026 $ \\
CA 2.2\,m         & Gunn $g$           & $0.1720 \pm 0.0054 $ & $0.0926  \pm 0.0024 $ & $89.9 \pm 1.4$ & $0.157  \pm 0.050 $ & $0.01458 \pm 0.00057 $ \\
CA 2.2\,m         & Gunn $r$           & $0.1692 \pm 0.0051 $ & $0.0959  \pm 0.0018 $ & $88.7 \pm 1.4$ & $0.1544 \pm 0.0046$ & $0.01480 \pm 0.00059$ \\
CA 2.2\,m         & Gunn $z$           & $0.1695 \pm 0.0059 $ & $0.0917  \pm 0.0016 $ & $88.3 \pm 1.5$ & $0.1552 \pm 0.0053$ & $0.01424 \pm 0.00058$ \\
INT               & Str\"{o}mgren $y$  & $0.1699 \pm 0.0052 $ & $0.0886  \pm 0.0016 $ & $87.9 \pm 1.2$ & $0.1560 \pm 0.0047$ & $0.01383 \pm 0.00056$ \\
Kuiper            & Cousins $I$        & $0.175  \pm 0.012  $ & $0.0871  \pm 0.0026 $ & $86.4 \pm 1.6$ & $0.161  \pm 0.012 $ & $0.0140  \pm 0.0010 $ \\
UDEM              & Cousins $I$        & $0.220  \pm 0.032  $ & $0.0836  \pm 0.0039 $ & $81.7 \pm 2.6$ & $0.203  \pm 0.030 $ & $0.0170  \pm 0.0021 $ \\
Adagio            & Cousins $V$        & $0.192  \pm 0.017  $ & $0.0956  \pm 0.0034 $ & $85.3 \pm 2.1$ & $0.175  \pm 0.015 $ & $0.0167  \pm 0.0018 $ \\
OHP               & Cousins $V$        & $0.172  \pm 0.031  $ & $0.0977  \pm 0.0088 $ & $87.4 \pm 3.9$ & $0.157  \pm 0.027 $ & $0.0153  \pm 0.0031 $ \\
Quarryview        & Red filter         & $0.169  \pm 0.011  $ & $0.0901  \pm 0.0024 $ & $87.5 \pm 2.1$ & $0.155  \pm 0.010 $ & $0.0140  \pm 0.0011 $ \\
Oversky           & Sloan $r^{\prime}$ & $0.171  \pm 0.016  $ & $0.0888  \pm 0.0033 $ & $86.9 \pm 2.6$ & $0.157  \pm 0.015 $ & $0.0140  \pm 0.0016 $ \\
KeplerCam         & Sloan $z$          & $0.1781 \pm 0.0081 $ & $0.0955  \pm 0.0013 $ & $86.9 \pm 1.3$ & $0.1626 \pm 0.0073$ & $0.01553 \pm 0.00084$ \\
\hline %
Final results     & & & $\mathbf{0.09208 \pm 0.00049}$ & $\mathbf{87.08 \pm 0.36}$ & $\mathbf{0.1590 \pm 0.0014}$ & $\mathbf{0.01468 \pm 0.00017}$ \\
\hline %
\citet{latham2009}  & & $0.1725_{-0.00048}^{+0.00094}$ & $0.0953  \pm 0.0009$ & $87.5_{-0.9}^{+1.9}$ & $0.1575_{-0.0042}^{+0.0084}$ & $0.01501_{-0.00054}^{+0.00095}$ \\
\citet{simpson2011} & & $0.1783_{-0.00063}^{+0.00060}$ & $0.09135 \pm 0.00089$ & $87.80_{-0.77}^{+0.75}$ & $0.1634_{-0.0056}^{+0.0053}$ & $0.01493_{-0.00066}^{+0.00063}$ \\
\hline %
\end{tabular}
\end{table*}
%

%%%%%%%%%%%%%%%%%%%%%%%%%%%%%%%%%%%%%%%%%%%%%%%%%%%%%%
\subsection{Datasets taken from literature}%
\label{sec_3.1}
%%%%%%%%%%%%%%%%%%%%%%%%%%%%%%%%%%%%%%%%%%%%%%%%%%%%%%
\citet{latham2009} reported three $z$-band transits (1451 points
in total), only one of which was observed in its entirety,
obtained with the 1.2\,m telescope and KeplerCam at the F.\ L.\
Whipple Observatory, US. We converted the timestamps into orbital
phase (using their ephemeris), sorted them and binned them into
154 points.

\citet{moutou2011} presented five datasets, two of them obtained
by the Adagio Association with an 82\,cm telescope ($V$ band; 720
points), one from the 1.20\,m telescope at Observatoire de
Haute-Provence ($V$ band; 256 points), one from a 35\,cm telescope
at the Oversky observatory ($R$ band; 569 points binned in 113
points), and one from a 32\,cm telescope at the Quarryview
Observatory ($R$ band, 157 points). The two Adagio datasets were
phased-binned into 144 points before analysis, to save computing
time. The Oversky data were also binned, into 113 points.

One transit was observed with the 36\,cm Universidad de Monterrey
Observatory (UDEM) telescope ($I$ band; 639 points). These data
were reported by \citet{todorov2012}. Finally, one very good light
curve was obtained by F.\ Hormuth using the CAHA 1.23\,m telescope
($R$ band; 176 points). These data are available from TRESCA web
site and were already studied by \citet{simpson2011}.

All of the light curves in this Section were fitted using {\sc
jktebop} in the same manner as our own data.

%%%%%%%%%%%%%%%%%%%%%%%%%%%%%%%%%%%%%%%%%%%%%%%%%%%%%%
\subsection{Final photometric parameters}%
\label{sec:finphot}
%%%%%%%%%%%%%%%%%%%%%%%%%%%%%%%%%%%%%%%%%%%%%%%%%%%%%%
All the light curves we fitted are shown in Figs.\,\ref{Fig_01}
and \,\ref{Fig_02}. The parameters of the fits are given in
Table\,\ref{Table_3}. In order to calculate the final photometric
parameters we took the weighted mean of the individual values and
uncertainties. This process highlighted the poor agreement between
the different datasets. The $\chi_{\nu}^2$ of the parameters of
the individual light curves with respect to the final weighted
means are $1.5$ for $r_{\rm A}$, $2.4$ for $i$ and $2.7$ for
$r_{\rm b}$. The worst agreement is found for $k$, where
$\chi_{\nu}^2=6.9$. Analagous situations have been found many
times in the course of the {\it Homogeneous Studies} project
\citep{southworth08,southworth09,southworth10,southworth11,southworth12},
but the agreement is rarely this poor.

Inspection of Fig.\,\ref{Fig_01} gives a clue to this problem. The
residuals of the fits to almost all of the available light curves,
including our own, exhibit systematic deviations from zero. This
correlated noise means the amount of information available in the
observational data is lower than suggested by the number of
datapoints and their uncertainties. We have already increased our
errorbars due to correlated noise using the $\beta$ approach
(Sect.\,\ref{sec_3}), but this was not enough to solve the
problem. Fortunately, we have 18 separate light curves which have
been analysed independently. Making the assumption that correlated
noise in data taken on different nights using different telescopes
is independent, we can account for it by inflating the errorbars
on the final photometric parameters by $\sqrt{\chi_{\nu}^2}$ to
give $\chi_{\nu}^2 = 1.0$ for each parameter. Because our results
are based on 18 different datasets, we are confident that our
analysis has yielded reliable parameters. The light curve
anomalies are discussed further in Sect.\,\ref{sec_5}.

%%%%%%%%%%%%%%%%%%%%%%%%%%%%%%%%%%%%%%%%%%%%%%%%%%%%%%
\subsection{Orbital period determination}
\label{sec_3.2}
%%%%%%%%%%%%%%%%%%%%%%%%%%%%%%%%%%%%%%%%%%%%%%%%%%%%%%

% Table 4
\begin{table*}
\caption{Transit mid-times of HAT-P-8 and their residuals. TRESCA
and AXA refer to the ``TRansiting ExoplanetS and CAndidates'' and
``Amateur Exoplanet Archive'' websites, respectively}
\label{Table_4} %
\centering %
\tiny
\begin{tabular}{lrrl}
% columns
\hline\hline
Time of minimum    & Cycle & Residual & Reference  \\
BJD(TDB)$-2400000$ & no.   & (JD)     &            \\
\hline %
$54437.67461 \pm 0.00044 $ &   0 &  0.00046 & \citet{latham2009}                 \\
$55034.4824  \pm 0.0031  $ & 194 & -0.00284 & Srdoc (AXA)                        \\
$55034.4881  \pm 0.0017  $ & 194 &  0.00286  & Ku\v{c}\'{a}kov\'{a} H. (TRESCA)  \\
$55046.7904  \pm 0.0008  $ & 198 & -0.00022 & Norby (AXA)                        \\
$55071.400   \pm 0.002   $ & 206 & -0.00138 & Ayiomamitis (AXA)                  \\
$55071.4017  \pm 0.0009  $ & 206 & -0.00032 & Srdoc (AXA)                        \\
$55071.4056  \pm 0.0012  $ & 206 &  0.00422 & Ku\v{c}\'{a}kov\'{a} H. (TRESCA)   \\
$55071.4132  \pm 0.0013  $ & 206 &  0.01182 & Trnka J. (TRESCA)                  \\
$55074.4752  \pm 0.0013  $ & 207 & -0.00253 & Srdoc (AXA)                        \\
$55123.7021  \pm 0.0018  $ & 223 &  0.00284 & Tieman B. (TRESCA)                 \\
$55126.77446 \pm 0.00079 $ & 224 & -0.00115 & This work (Kuiper 155\,cm)         \\
$55148.3140  \pm 0.0019  $ & 231 &  0.00397 & This work (Loiano 152\,cm)         \\
$55409.7992  \pm 0.0012  $ & 316 & -0.00022 & \cite{todorov2012} (Udem 36\,cm)   \\
$55434.4108  \pm 0.0023  $ & 324 &  0.00061 & \cite{moutou2011} (OHP 120\,cm)    \\
$55437.48610 \pm 0.00043 $ & 325 & -0.00043 & This work (INT 250\,cm)            \\
$55437.4834  \pm 0.0011  $ & 325 & -0.00313 & \cite{moutou2011} (Oversky 35\,cm) \\
$55437.48434 \pm 0.00088 $ & 325 & -0.00219 & Hormuth F. (CA 123\,cm)            \\
$55437.4895  \pm 0.0010  $ & 325 &  0.00297 & Ruocco N. (TRESCA)                 \\
$55440.5646  \pm 0.0012  $ & 326 &  0.00171 & Marino G. (TRESCA)                 \\
$55449.79074 \pm 0.00099 $ & 329 & -0.00118 & \cite{moutou2011} (Hose 32cm)      \\
$55452.86716 \pm 0.00079 $ & 330 & -0.00110 & Hose K. (TRESCA)                   \\
$55797.4152  \pm 0.0018  $ & 442 & -0.00379 & D\v{r}ev\v{e}n\'{y} R., Kuchtak B. (TRESCA) \\
$55797.4186  \pm 0.0015  $ & 442 & -0.00039 & Br\'{a}t L. (TRESCA)               \\
$55800.4916  \pm 0.0013  $ & 443 & -0.00374 & Br\'{a}t L. (TRESCA)               \\
$55800.4942  \pm 0.0011  $ & 443 & -0.00114 & Trnka J. (TRESCA)                  \\
$55800.4968  \pm 0.0025  $ & 443 &  0.00146 & Zibar M. (TRESCA)                  \\
$55800.4996  \pm 0.0010  $ & 443 &  0.00426 & Marek P. (TRESCA)                  \\
$55812.80021 \pm 0.00085 $ & 447 & -0.00051 & Shadic S. (TRESCA)                 \\
$55834.3372  \pm 0.0019  $ & 454 &  0.00206 & Br\'{a}t L. (TRESCA)               \\
$55837.4136  \pm 0.0025  $ & 455 &  0.00211 & Trnka J. (TRESCA)                  \\
$55840.48811 \pm 0.00063 $ & 456 &  0.00028 & This work (CA 123\,cm)             \\
$55840.48845 \pm 0.00049 $ & 456 &  0.00062 & This work (Loiano 152\,cm)         \\
$55886.6345  \pm 0.0021  $ & 471 &  0.00148 & Dvorak S. (TRESCA)                 \\
$56166.5776  \pm 0.0011  $ & 562 & -0.00289 & This work (CA 220\,cm BUSCA-$u$)   \\
$56166.58178 \pm 0.00057 $ & 562 &  0.00129 & This work (CA 220\,cm BUSCA-$g$)         \\
$56166.58162 \pm 0.00051 $ & 562 &  0.00113 & This work (CA 220\,cm BUSCA-$r$)         \\
$56166.57924 \pm 0.00054 $ & 562 & -0.00124 & This work (CA 220\,cm BUSCA-$z$)         \\
$56175.81123 \pm 0.00059 $ & 565 &  0.00170 & Hose K. (TRESCA)                   \\
$56203.49662 \pm 0.00031 $ & 574 & -0.00002 & This work (Loiano 152\,cm)         \\
$56203.49682 \pm 0.00083 $ & 574 &  0.00018 & This work (CA 123\,cm)             \\
\hline %
\end{tabular}
\end{table*}

The transit time for each dataset was obtained using {\sc
jktebop}, and uncertainties were estimated by Monte Carlo
simulations. In the determination of the orbital period of the
HAT-P-8 system, we also considered 23 timings measured by amateur
astronomers and available on the ETD\footnote{The Exoplanet
Transit Database (ETD) website can be found at
http://var2.astro.cz/ETD} website (see Table\,\ref{Table_4}).

\begin{figure*}%
\centering
\includegraphics[width=16.cm]{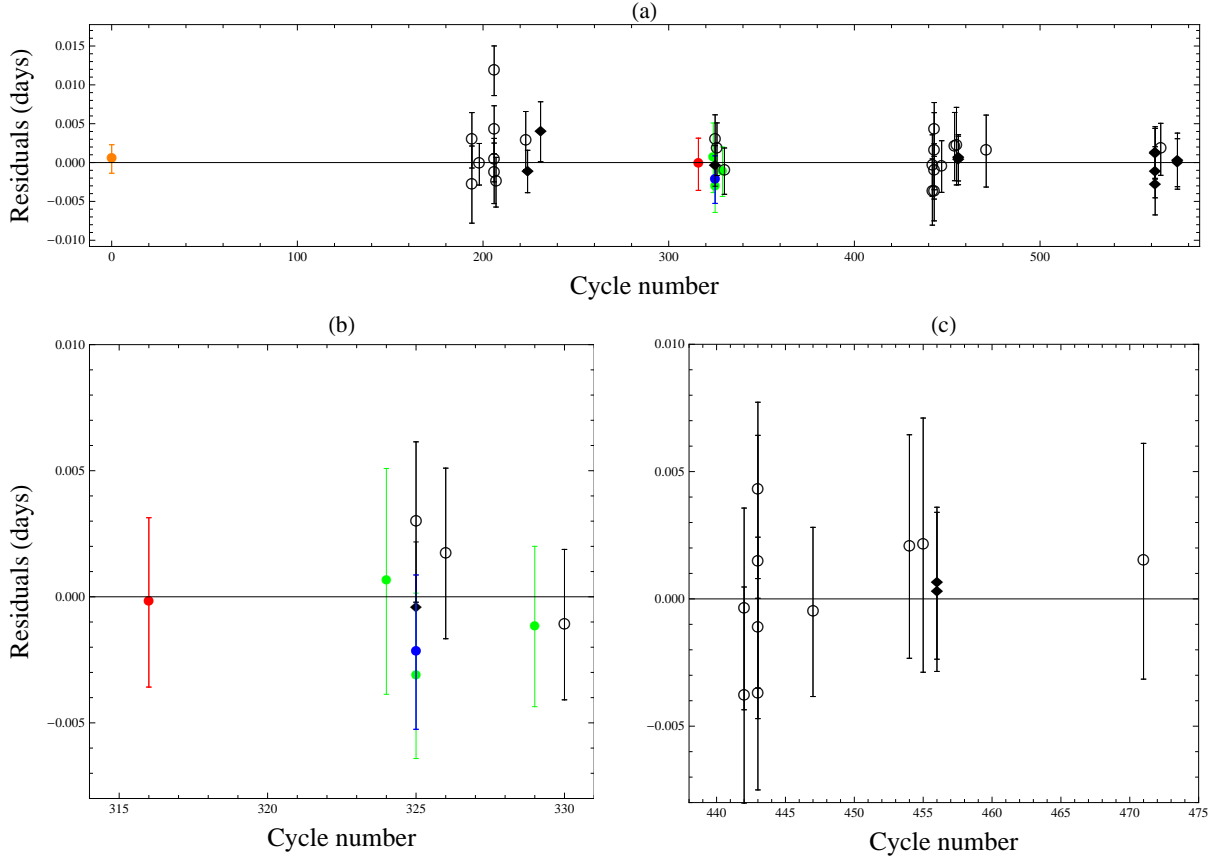}
\caption{Panel (a): plot of the residuals of the timings of
mid-transit of HAT-P-8\,b versus a linear ephemeris. The timings
in black are from this work, orange from \citet{latham2009}, green
from \citet{moutou2011}, red from \citet{todorov2012}, and blue
from Hormuth (TRESCA). The other timings obtained by amateur
astronomers are plotted using open circles. Panels (b) and (c) are
zooms in to the two best sampled regions.} %
\label{Fig_03} %
\end{figure*}

We excluded from the analysis the incomplete ETD light curves and
those with a Data Quality index higher than 3. All timings were
placed on BJD(TDB) time system. The resulting measurements of
transit midpoints were fitted with a straight line to obtain a new
orbital ephemeris:
{\small
\begin{equation}
T_{0} = \mathrm{BJD(TDB)} 2\,454\,437.6742(14) + 3.0763458(24)\,E,
\end{equation}
}
where $E$ is the number of orbital cycles after the reference
epoch (which we take to be the midpoint of the first transit
observed by \citealt{latham2009}) and quantities in brackets
denote the uncertainty in the final digit of the preceding number.
The fit has $\chi_{\nu}^2=5.30$, and the uncertainties given above
have been increased to account for this. The large $\chi_{\nu}^2$
indicates that the uncertainties in the various $T_0$ measurements
are too small, most probably due to the systematic differences
between the light curves and their best fits as noted in
Sect.\,\ref{sec:finphot}. A plot of the residuals around the fit
is shown in Fig.\,\ref{Fig_03} and does not indicate any clear
systematic deviation from the predicted transit times. We
therefore take the conservative option of not interpreting the
large $\chi_{\nu}^2$ as a suggestion of transit timing variations.

% Figure 03
%

%%%%%%%%%%%%%%%%%%%%%%%%%%%%%%%%%%%%%%%%%%%%%%%%%%%%%%
\section{The physical properties of HAT-P-8}
\label{sec_4}
%%%%%%%%%%%%%%%%%%%%%%%%%%%%%%%%%%%%%%%%%%%%%%%%%%%%%%
Following \citet{southworth09}, we estimated the physical
properties of the HAT-P-8 system from the photometric parameters
measured in Sect.\,\ref{sec_3}, published spectroscopic results,
and theoretical stellar models. The orbital eccentricity, velocity
amplitude and metallicity of the star, ($e=0$,
$K_{\mathrm{A}}=153.1 \pm 3.9$, [Fe/H]$=+0.01 \pm 0.08$) were
taken from \citet{latham2009}, and for the parent-star effective
temperature ($T_{\mathrm{eff}}=6130 \pm 80$) we used that measured
by \citet{knutson2010}.

An initial estimate of the velocity amplitude of the planet
($K_{\rm b}$) was iteratively refined by calculating the system
properties using standard formulae (e.g.\ \citealp{hilditch2001}),
and comparing the observed $T_{\mathrm{eff}}$ and $r_{\mathrm{A}}$
with values of $T_{\mathrm{eff}}$ and $R_{\mathrm{A}}/a$ predicted
by theoretical models for the calculated mass of the star. This
calculation was performed over a grid of ages and for five
different sets of stellar models (see \citealt{southworth10}).
Statistical errors were propagated by a perturbation analysis. The
resulting estimates of the physical properties are given in
Table\,\ref{Table_5}. The final set of physical properties was
obtained by taking the unweighted mean of the five sets of values
found from the different stellar models, which also allowed us to
obtain an estimate of the systematic errors inherent in the use of
stellar theory. The results of this process are given in
Table\,\ref{Table_6}.

Finally, following the method delineated by \citet{enoch2010} and
improved by \citet{southworth11}, we used empirical measurements
of stars in detached eclipsing binary (dEB) systems to calibrate
the parent star of the HAT-P-8 system. This allowed us to measure
the physical properties of the system without using stellar
models, thus avoiding the dependence on stellar theory. These
results are also given in Table\,\ref{Table_5}.

% Table 5
\begin{table*}
\caption{Derived physical properties of the HAT-P-8 planetary system.}%
\label{Table_5} %
\centering %
\tiny
\begin{tabular}{lcccccc}
% columns
\hline\hline
& This work & This work & This work & This work & This work & This work  \\
& (dEB constraint) & ({\sf Claret} models) & ({\sf Y$^2$} models) & ({\sf Teramo} models) & ({\sf VRSS} models) & ({\sf DSEP} models) \\
\hline %
$K_{\mathrm{b}}$ (km\,s$^{-1}$)             & $159.1   \pm 3.9$     & $154.8   \pm 2.3$     & $156.7   \pm 1.9$     & $153.4   \pm 1.7$     & $154.2   \pm 2.6$     & $155.4   \pm 2.13$     \\
$M_{\mathrm{A}}$ ($M_{\sun}$)               & $1.292   \pm 0.094$   & $1.189   \pm 0.054$   & $1.234   \pm 0.044$   & $1.158   \pm 0.037$   & $1.175   \pm 0.061$   & $1.202   \pm 0.053$   \\
$R_{\mathrm{A}}$ ($R_{\sun}$)               & $1.521   \pm 0.041$   & $1.480   \pm 0.032$   & $1.470   \pm 0.026$   & $1.467   \pm 0.023$   & $1.474   \pm 0.031$   & $1.485   \pm 0.026$   \\
$\log g_{\mathrm{A}}$ (cgs)                 & $4.1852  \pm 0.0137$  & $4.1733  \pm 0.0088$  & $4.1954  \pm 0.0103$  & $4.1694  \pm 0.0105$  & $4.1715  \pm 0.0106$  & $4.1748  \pm 0.0115$   \\
$M_{\mathrm{b}}$ ($M_{\mathrm{jup}}$)       & $1.345   \pm 0.071$   & $1.273   \pm 0.048$   & $1.305   \pm 0.041$   & $1.251   \pm 0.037$   & $1.263   \pm 0.053$   & $1.282   \pm 0.047$   \\
$R_{\mathrm{b}}$ ($R_{\mathrm{jup}}$)       & $1.357   \pm 0.043$   & $1.320   \pm 0.034$   & $1.337   \pm 0.032$   & $1.308   \pm 0.030$   & $1.315   \pm 0.037$   & $1.325   \pm 0.034$   \\
$\rho_{\mathrm{b}}$ ($\rho_{\mathrm{jup}}$) & $0.503   \pm 0.034$   & $0.517   \pm 0.034$   & $0.511   \pm 0.033$   & $0.522   \pm 0.034$   & $0.519   \pm 0.035$   & $0.516   \pm 0.034$   \\
$\Theta$                                    & $0.0691  \pm 0.0026$  & $0.0711  \pm 0.0023$  & $0.0702  \pm 0.0022$  & $0.0717  \pm 0.0022$  & $0.0714  \pm 0.0025$  & $0.0708  \pm 0.0023$  \\
$a$ (AU)                                    & $0.04510 \pm 0.00109$ & $0.04387 \pm 0.00066$ & $0.04442 \pm 0.00053$ & $0.04348 \pm 0.00047$ & $0.04370 \pm 0.00075$ & $0.04403 \pm 0.00065$ \\
Age                                         & $-$                   & $4.7_{-0.5}^{+1.4}$   & $3.5_{-0.7}^{+0.4}$   & $4.7_{-0.5}^{+1.0}$   & $4.2_{-0.4}^{+1.5}$   & $4.2_{-1.4}^{+0.5}$   \\
\hline
\end{tabular}
\end{table*}

% Table 6
\begin{table*}
\caption{Final physical properties of the HAT-P-8 system, compared
with results from the literature.
Where two errorbars are given, the first refers to the statistical uncertainties and the second to the systematic errors.}%
\label{Table_6} %
\centering
\begin{tabular}{lcccc}
% columns
\hline\hline
& This work (final) & \citet{latham2009} & \citet{moutou2011} & Latham (private comm.) \\
\hline %
$M_{\mathrm{A}}$ ($M_{\sun}$)               & $1.192   \pm 0.061   \pm 0.043$   & $1.28   \pm 0.04$      & $-$                 & $1.28_{-0.06}^{+0.04}$       \\
$R_{\mathrm{A}}$ ($R_{\sun}$)               & $1.475   \pm 0.032   \pm 0.010$   & $1.58^{+0.08}_{-0.06}$ & $-$                 & $1.57  \pm 0.07$             \\
$\log g_{\mathrm{A}}$ (cgs)                 & $4.177   \pm 0.011   \pm 0.019$   & $4.15   \pm 0.03$      & $-$                 & $4.15  \pm 0.03$             \\
$\rho_{\mathrm{A}}$ ($\rho_{\sun}$)         & $0.371   \pm 0.013   \pm 0.018$   & $-$                    & $-$                 & $0.46  \pm 0.05$             \\
$M_{\mathrm{b}}$ ($M_{\mathrm{jup}}$)       & $1.275   \pm 0.053   \pm 0.030$   & $1.52^{+0.18}_{-0.16}$ & $1.34 \pm 0.05$     & $1.38  \pm 0.05$             \\
$R_{\mathrm{b}}$ ($R_{\mathrm{jup}}$)       & $1.321   \pm 0.037   \pm 0.016$   & $1.50^{+0.08}_{-0.06}$ & $-$                 & $1.40  \pm 0.08$             \\
$g_{\mathrm{b}}$ ($\mathrm{ms^{-1}}$)       & $18.11   \pm 0.82$                & $16.98  \pm 1.17$      & $-$                 & $1.38  \pm 0.05$             \\
$\rho_{\mathrm{b}}$ ($\rho_{\mathrm{jup}}$) & $0.517   \pm 0.035   \pm 0.006$   & $0.568  \pm 0048$      & $-$                 & $0.62  \pm 0.09$             \\
$T_{\mathrm{eq}}$ ($\mathrm{K}$)            & $1713    \pm 24      \pm 13   $   & $1700   \pm 35$        & $-$                 & $1771  \pm 39  $             \\
$\Theta$                                    & $0.0710  \pm 0.0025  \pm 0.0008$  & $0.061  \pm 0.003$     & $-$                 & $0.069 \pm 0.004$            \\
$a$ (AU)                                    & $0.04390 \pm 0.00075 \pm 0.00052$ & $0.0487 \pm 0.0026$    & $0.0449 \pm 0.0007$ & $0.0450_{-0.0007}^{+0.0004}$ \\
Age (Gyr)                                   & $4.3_{-1.4\,-0.5}^{+1.5\,+0.8}$   & $3.4    \pm 1.0$       & $-$                 & $3.3_{-0.3}^{+0.7}$          \\
\hline
\end{tabular}
\end{table*}

\begin{figure}%
\centering
\includegraphics[width=9.cm]{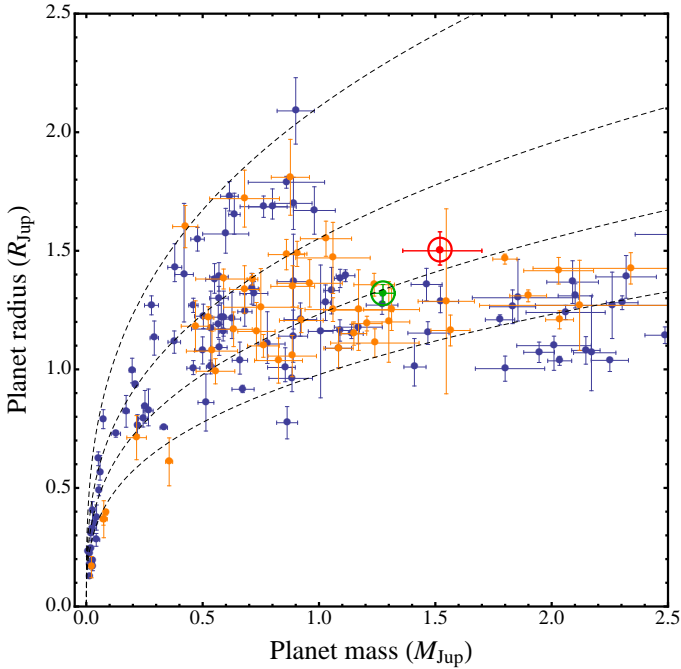}
\caption{Plot of the masses and radii of the known TEPs. The
orange symbols denote values from the {\it Homogeneous Studies}
project and the blue symbols results for the other known TEPs.
HAT-P-8\,b is shown in red \citep{latham2009} and green (this
work). Dotted lines show where density is (from bottom to top)
1.0, 0.5, 0.25 and 0.1 $\rho_{\mathrm{Jup}}$.} \label{Fig_04}
\end{figure}

Table\,\ref{Table_6} also contains a comparison between our own
results and those found by \citet{latham2009}. We find smaller
masses and radii for both the planet and the host star. This is
surprising because the values of the fractional radius of the
star, which furnishes the vital constraint on the stellar density,
are very similar between the two studies (we find $r_{\rm A} =
0.1590 \pm 0.0014$ versus $r_{\rm A} = 0.1575^{+0.0041}_{-0.0089}$
for \citealt{latham2009}). One small difference is that we adopted
$T_{\mathrm{eff}} = 6130 \pm 80$\,K \citep{knutson2010} as opposed
to the value of $6200 \pm80$\,K used by \citet{latham2009}. Our
revised physical properties move HAT-P-8\,b into a region of
parameter space which is comparatively well-supplied with
transiting planets. Fig.\,\ref{Fig_04} shows the change in
position in the planet mass--radius plot.

% Figure 04

We contacted D.\ Latham in order to check the veracity of the
final values of the parameters that they reported in their
discovery paper. D.\ Latham confirmed that some of these results
are indeed correct, and has kindly supplied revised values. These
are from a re-analysis carried out by J.\ Hartman on a slight
updated dataset, and are given in the final column of
Table\,\ref{Table_6}. They correspond to a smaller planetary mass
and radius ($M_{\mathrm{b}}=1.38\,M_{\mathrm{jup}}$,
$R_{\mathrm{b}}=1.40\,R_{\mathrm{jup}}$) than given in
\citet{latham2009}, in agreement with our own findings.

%%%%%%%%%%%%%%%%%%%%%%%%%%%%%%%%%%%%%%%%%%%%%%%%%%%%%%
\section{Variation of planetary radius with wavelength}
\label{sec_5}
%%%%%%%%%%%%%%%%%%%%%%%%%%%%%%%%%%%%%%%%%%%%%%%%%%%%%%
One of the factors that plays a principal role in determining the
atmospheric properties of hot-Jupiter planets is the amount of
stellar flux incident on the planet's surface. Variations in this
irradiation cause different planets to have different atmospheric
chemical mixing ratios and atmospheric opacities.  This could
potentially lead to divide into classes of hot Jupiters.  An
initial suggestion was that of \citet{fortney2008} to
distinguished pM- and pL-class planets, depending on the presence
of strong absorbers such as gaseous titanium oxide (TiO) and
vanadium oxide (VO) in their atmospheres. By observing a planetary
transit at different wavelengths, it is possible to detect a
variation in the value of the radius measured as a function of the
wavelength, and probe chemistry and wavelength dependent opacity
at the planet's terminator. This dependence on the wavelength can
be therefore used to probe the atmospheric composition of TEPs,
being a complementary method to the observations of secondary
eclipses.

\begin{figure}%
\centering
\includegraphics[width=9.cm]{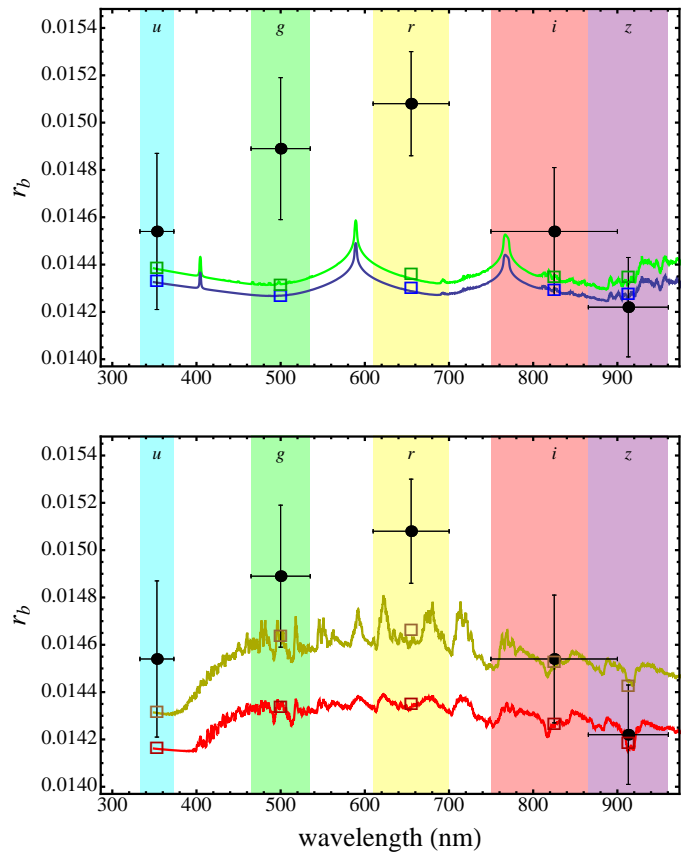}
\caption{Variation of the fractional planetary radius
$r_{\mathrm{b}}=R_{\mathrm{b}}/a$ with wavelength. The points
shown in the plot are from the Calar Alto 2.2\,m ($u$, $g$, $r$
and $z$ bands) and Loiano ($i$-band) telescopes. The vertical bars
represent the errors in the measurements and the horizontal bars
show the full widths at half maximum transmission of the passbands
used. The experimental points are compared with four models. These
use planet-wide average pressure-temperature profiles (in blue and
red) or warmer day-side average profiles (in green and yellow).
Synthetic spectra in the top panel do not include TiO and VO
opacity, while spectra in the bottom do, based on equilibrium
chemistry. Coloured squares represent band-averaged model radii
over the bandpasses of the observations.} %
\label{Fig_05}
\end{figure}

As an additional possibility offered by the BUSCA data, we made an
attempt to investigate possible variations of the radius of
HAT-P-8\,b in different optical passbands. Receiving from its
parent star an incident flux of $2.22\pm0.20\times10^9$
erg\,s$^{-1}$\,cm$^{-2}$ (Latham, private communication),
HAT-P-8\,b should belong to the pM class of planets. The
theoretical models of \citet{fortney2010} predict that its radius
should be few percent lower at $350$--$400$\,nm and
$800$--$950$\,nm versus $500$--$750$\,nm. Following the strategy
used by \citet{southworth2012b}, we fitted the BUSCA light curves
with all parameters fixed to the final values reported in
Table\,\ref{Table_4}, with the exception of $k$ and the LD
coefficients. The errors were estimated by a residual-permutation
algorithm. The results are exhibited in Fig.\,\ref{Fig_05}, where
the points show the data, the vertical bars represent the relative
errors in the measurements and the horizontal bars show the full
widths at half maximum transmission of the passbands used. As is
apparent from Fig.\,\ref{Fig_02}, the BUSCA-$z$ light curve shows
some structure in its residuals from phase 0 to 0.02, which
suggests that the $z$-band data suffer correlated noise. The
result coming from this band should be therefore considered with
caution. The optical region not covered by the BUSCA data was
investigated by using the $i$-band light curve from Loiano, which
is the best one that we have at this band (see
Fig.\,\ref{Fig_01}).

A variation of $r_{\mathrm{b}}$ along the five passbands is
clearly visible in Fig.\,\ref{Fig_05}. We investigated this
variation with the help of model atmospheres. We used detailed
non-gray atmosphere codes to model the temperature structure and
transmission spectrum of the planet. We computed 1D model
atmosphere of HAT-P-8\,b, using the atmosphere code described in
\citet{fortney2005,fortney2008}. Pressure-temperature profiles are
derived that either include or exclude the opacity of TiO and VO
molecules. The fully non-gray model uses the chemical equilibrium
abundances of \citet{Lodders02} and the opacity database described
in \citet{fr08}. The atmospheric pressure-temperature profiles
simulate planet-wide average conditions or day-side average
conditions. We computed the transmission spectrum of the models
using the methods described in \citet{fortney2010} and
\citet{shabram2011}.

Excluding TiO and VO from the opacity calculation, in the upper
panel of Fig.5 we compare the transmissions spectrum of the 1D
planet-wide average profile (in blue) with a warmer day-side
average model (in green), to experimental data. In comparison, the
bottom panel of Fig. 5 shows analogous calculations in red and
yellow respectively, which include both the TiO and VO species.
Coloured boxes indicate the predicted values for these models
integrated over the bandpasses of the observations. Models without
TiO and VO have optical transmission spectra that are dominated by
Rayleigh scattering in the blue, and pressure-broadened neutral
atomic lines of Na an K at 589 nm and 770 nm, respectively. Models
with TiO and VO absorption show significant optical absorption (a
much large transit radius), that broadly peaks around 700 nm, with
a sharp falloff in the blue, and a shallower falloff in the red.

Comparing the panels of Fig.\,\ref{Fig_05}, it is readily apparent
that the model that gives the best match to the data is the one
with TiO and VO opacity in the atmosphere of HAT-P-8\,b. The
increased optical radius is somewhat larger than the model
prediction. Taken at face value, the observations are in general
agreement with the \citet{fortney2008}'s hot-Jupiter
classification based on stellar irradiation.  We suggest that
HAT-P-8\,b should be an important target for followup studies to
confirm or refute these suggestive observations.  The clear
detection of TiO/VO, or other strong optical absorbers, would be
an important step in characterizing hot Jupiter atmospheres, as
such absorbers are thought to cause temperature inversions in
these atmospheres (e.g. \citealp{fortney2008, burrows2008}).
Previously, \citet{desert2008} suggested a detection of TiO and VO
at strongly depleted levels in the red optical spectrum of HD
209845\,b, using Hubble STIS.

% Figure 05

%%%%%%%%%%%%%%%%%%%%%%%%%%%%%%%%%%%%%%%%%%%%%%%%%%%%%%
\section{Light-curve anomalies}
\label{sec_6}
%%%%%%%%%%%%%%%%%%%%%%%%%%%%%%%%%%%%%%%%%%%%%%%%%%%%%%

% Figure 06
\begin{figure}%
\centering
\includegraphics[width=9.cm]{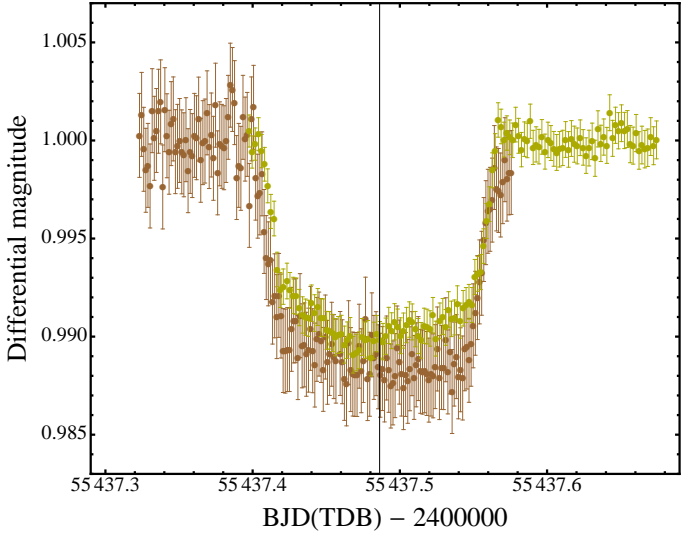}
\caption{An example of the consistency of follow-up observations
made of the same transit on the night of 2010/08/28 by different
telescopes. This is an example of a ``regular'' transit. Lighter
yellow dots are for the data taken at the INT (this work), whereas
the darker brown ones for those taken by F.\ Hormuth (ETD) at the
CAHA 1.23\,m. The agreement between the two datasets is not very
good. The difference in depth between the two datasets is partly
due to different filters used, Str\"{o}mgren $y$ and Johnson $R$,
respectively. The vertical line represents the expected transit
minimum time.}%
\label{Fig_06}
\end{figure}

While one of our follow-up light curves, obtained at the INT,
displays a regular transit shape, the others all show anomalies of
a similar shape. They display an asymmetry with respect the line
of minimum transit time. It is important to clarify if the
anomalies that we detected have an astrophysical nature or are of
instrumental or environmental origin. One way to check this is to
have independent measures of the same transit event, obtained
using multiple telescopes located at different sites. In this way,
if both the telescopes reported the same anomaly, it less likely
that they are caused by instrumental or Earth-atmosphere effects.
This already happened inadvertently in the case of several
follow-up observations of WASP-33 carried out by amateur
astronomers and reported in \citet{kovacs2012}. It also occurred
by chance for the transit observations of HAT-P-8 on the night of
2010/08/28, which was observed by ourselves at the INT and
contemporaneously at CAHA by F.\ Hormuth. Fig.\,\ref{Fig_06} shows
the two light curves in the same plot; we note that both have a
regular transit shape but disagree over the transit depth. Some
fraction of this disagreement is due to the different LD
characteristics, as the INT data were obtained with a bluer filter
than the CAHA data.

\begin{figure}%
\centering
\includegraphics[width=9.cm]{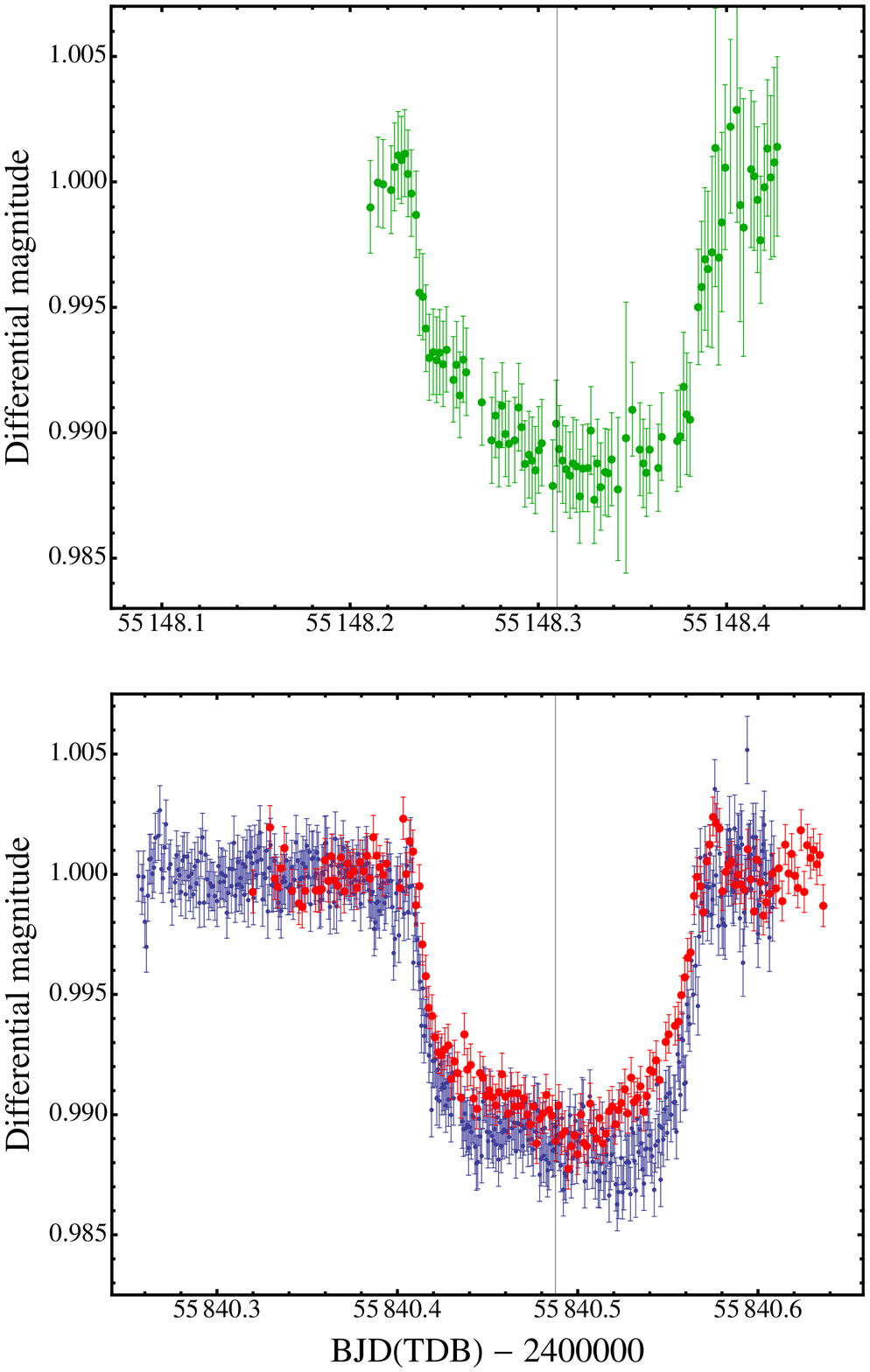}
\caption{Upper panel: an example of a discrepant transit. The
green points are for the data taken at the Cassini telescope (this
work) through a Gunn $r$ filter. Lower panel: an example of the
consistency of followup observations made on the same date of
2011/10/05 by different observers. Lighter blue dots are for the
data taken at the Cassini telescope (this work), whereas the red
ones for those taken at CAHA 1.23\,m (this work). Some of the
difference in depth between the two datasets is due to different
filters used, Gunn $i$ and Johnson $R$, respectively.
The vertical line represents the expected transit minimum time.}%
\label{Fig_07}
\end{figure}

In 2011, we aimed to observe transits in the HAT-P-8 system from
two different observatories and this goal was successfully
achieved on the night of 2011/10/05 when a transit was
simultaneously observed using the 1.52\,m Cassini and CAHA 1.23\,m
telescopes. The datasets show partial but not complete agreement
about the transit shape anomalies, as well as slight differences
due to the different LD in the $i$ and $R$ passbands. Details of
the two light curves are displayed in the lower panel of
Fig.\,\ref{Fig_07}, while the upper panel shows another transit
(2009, Loiano) which presents a similar anomaly. The anomalies
cannot be removed by choosing different comparison stars for the
differential photometry, but are not completely consistent between
different datasets for the same transit. We have therefore treated
them as correlated noise in our analysis (see Sect.\,\ref{sec_3}).

% Figure 07

If the recurrent anomalies that we noticed in the HAT-P-8 light
curves are not of instrumental origin, is there a reasonable
astrophysical explanation for these features? We can easily
exclude a gravity darkening effect (e.g.\ \citealt{barnes2009})
because the rotation rate of the star is low
(\citealp{latham2009,moutou2011}). The presence of a moon is an
unlikely possibility, and in any case is difficult to constrain
with so few observations.

Another possibility is that the anomalies are due to single or a
belt of stellar spots on the photosphere of the parent star.
\citet{knutson2010} estimated that HAT-P-8 has an activity index
of $\log(R_{HK}^{\prime})= -4.985$, which indicates that the star
has a moderate chromospheric activity. This is related to the
star's magnetic structure and therefore to the presence of
photospheric features, such as spots, that modulate luminosity. It
is thus possible that, during the transit, dark starspots are
occulted by the planet. Similar features were indeed already
observed in several cases, e.g.\ HD\,189733 \citep{pont2007},
TrES-1 \citep{rabus2009}, Kepler-17 \citep{desert2011}, HAT-P-11
\citep{sanchis2011}, and WASP-19 \citep{tregloan2012}. One
argument against this is the relatively high $T_{\rm eff}$ of the
star, $6130 \pm 80$\,K, which makes starspots less likely to
occur. Moreover, spots would only affect small parts of the light
curve, while here both light curves show trends for the
\textit{full} duration of the transit.

Another possible explanation is that the faint M-dwarf companion
of HAT-P-8 is a \emph{flare star}, which could emit giant flares
bright enough to significantly affect the light curves. However,
the magnitude difference between the two objects ($\Delta
i^{\prime}=7.34 \pm 0.10$, \citealp{bergfors2012}), is such that
this star would have to generate a superflare in order to make a
noticeable difference. One more explanation for this behaviour is
that a significant fraction of the comparison stars display
intrinsic variability which is insufficient to detect in
individual stars but, when combined, is enough to modify the
transit shape in the HAT-P-8 system.

Having exhausted all possibilities, we conclude that the apparent
transit shape \textit{distortion} shown in Fig.\,\ref{Fig_07}
could be caused by differential color extinction, pathological
variability in the comparison stars, or other low-frequency noise
of atmospheric or astrophysical origin. Further investigations of
these hypotheses would, inevitably, require substantial further
data and is beyond the scope of the current work.

%%%%%%%%%%%%%%%%%%%%%%%%%%%%%%%%%%%%%%%%%%%%%%%%%%%%%%
\section{Summary and Conclusions}
\label{sec_7}
%%%%%%%%%%%%%%%%%%%%%%%%%%%%%%%%%%%%%%%%%%%%%%%%%%%%%%

We have reported observations of six transits of the HAT-P-8
system performed using five different medium-class telescopes, for
a total of eleven new light curves. All but one of these transits
were obtained using the defocussed-photometry technique, achieving
a photometric precision of $0.47$--$1.7$\,mmag per observation.
Four of them were observed on the same nights from two different
telescopes. In one of these two nights, both the light curves show
an anomaly which is probably caused by systematic noise of
atmospheric or astrophysical origin. Another transit was
simultaneously observed through four optical passbands in a
wavelength range between 330 and 960 nm.

We modelled our new and seven published datasets using the {\sc
jktebop} code, and used the results to determine the physical
properties of the planet and its host star. Compared to the
discovery paper \citep{latham2009}, we find a significantly
smaller radius and mass for HAT-P-8\,b. We obtain $R_{\rm b} =
1.321 \pm 0.037 \, R_{\mathrm{Jup}}$ versus $1.50_{-0.06}^{+0.08}
\, R_{\mathrm{Jup}}$, and $M_{\rm b} = 1.275 \pm 0.053 \,
M_{\mathrm{Jup}}$ versus $1.52_{-0.16}^{+0.18} \,
M_{\mathrm{Jup}}$. The theoretical radius calculated by
\citet{fortney2007} for a core-free planet at age 4.5\,Gyr and
distance 0.045\,AU is $1.107$--$1.108 \, R_{\mathrm{Jup}}$ for a
planet of mass in the range $1.0$--$1.46 \, M_{\mathrm{Jup}}$,
which is still significantly smaller than the radius we find.

\citet{latham2009} found that HAT-P-8\,b was a comparatively
highly inflated planet. Instead, our results place it firmly in a
well-populated part of the mass--radius diagram, removing its
outlier status. Whilst the existing transit light curves of the
HAT-P-8 system all show some systematic deviations from the best
fits found using simple geometric models, the large number of
available datasets means that our overall results are reliable.
HAT-P-8 is another system where extensive follow-up photometry has
been necessary to determine robust physical properties for a
planetary system.

Finally, thanks to the ability of BUSCA to measure stellar flux
simultaneously through different filters, covering a quite large
range of optical window, we probed the composition of the
atmosphere of HAT-P-8\,b by investigated how vary its radius in
these wavelengths. In fact, the presence of strong optical
absorbers in the atmosphere of the planet should produce larger
transit radius at optical wavelengths than in the near UV or near
infrared. We measured a variation of the radius of HAT-P-8\,b
along five passbands, corresponding to a wavelength coverage of
$330-960$ nm. This result was then theoretically investigated by
using several synthetic spectra based on isothermal model
atmospheres in chemical equilibrium. The comparison between the
models and the experimental points suggests the presence of
molecular gas that strongly absorbs in the optical, potentially
composed of TiO and VO gases, in the atmosphere of HAT-P-8\,b.

\begin{acknowledgements}
Based on observations collected at the Centro Astron\'{o}mico
Hispano Alem\'{a}n (CAHA) at Calar Alto, Spain, operated jointly
by the Max-Planck Institut f\"{u}r Astronomie and the Instituto de
Astrof\'{i}sica de Andaluc\'{i}a (CSIC); observations obtained
with the 1.52\,m Cassini telescope at OAB Loiano Observatory,
Italy; data collected at the Isaac Newton Telescope, operated on
the island of La Palma, by the Isaac Newton Group in the Spanish
Observatorio del Roque de los Muchachos. The reduced light curves
presented in this work will be made available at the CDS
(http://cdsweb.u-strasbg.fr/). We thank Kamen Todorov and Rodrigo
F. D\'{i}az for supplying photometric data. JS acknowledges
financial support from STFC in the form of an Advanced Fellowship.
We thank Ulli Thiele and Roberto Gualandi for their technical
assistance at the CA 2.2\,m telescope and Cassini telescope,
respectively. We thank the anonymous referee for their useful
criticisms and suggestions that helped us to improve the quality
of the present paper. The following internet-based resources were
used in research for this paper: the ESO Digitized Sky Survey; the
NASA Astrophysics Data System; the SIMBAD data base operated at
CDS, Strasbourg, France; and the arXiv scientific paper preprint
service operated by Cornell University. LM thanks the HESS site in
Namibia for the kind hospitality.
\end{acknowledgements}

\bibliographystyle{aa} % style aa.bst

\end{document}